\documentclass[floatfix
]{revtex4}
\usepackage{graphicx}
\usepackage{dcolumn}
\usepackage{bm}
\usepackage{amssymb}
\usepackage{times}
\usepackage{amsmath}
\usepackage{amsfonts}
\usepackage{mathrsfs,amssymb}
\usepackage{setspace}
\usepackage{epsfig}
\usepackage{latexsym}
\usepackage{bbm}
\usepackage{subfigure}
\usepackage{float}
\usepackage{flafter}
\usepackage{bm}
\usepackage{epstopdf}

\setcitestyle{numbers,round}

\begin{document}

\title{Nature of strong hole pairing in doped Mott antiferromagnets}
\author{Zheng Zhu, Hong-Chen Jiang, D. N. Sheng $\&$ Zheng-Yu Weng}

\begin{abstract}
\textbf{Cooper pairing instability in a Fermi liquid is well understood by the BCS theory, but pairing mechanism for doped Mott insulators still remains elusive. Previously it has been shown by density matrix renormalization group (DMRG) method that a single doped hole is always self-localized due to the quantum destructive interference of the phase string signs hidden in the $t$-$J$ ladders. Here we report a DMRG investigation of hole binding in the same model, where a novel pairing-glue scheme beyond the BCS realm is discovered. Specifically, we show that, in addition to spin pairing due to superexchange interaction, the strong frustration of the phase string signs on the kinetic energy gets effectively removed by pairing the charges, which results in strong binding of two holes. By contrast, if the phase string signs are ``switched off'' artificially, the pairing strength diminishes significantly even if the superexchange coupling remains the same. In the latter, unpaired holes behave like coherent quasiparticles with pairing drastically weakened, whose sole origin may be attributed to the resonating-valence-bond (RVB) pairing of spins. Such non-BCS pairing mechanism is therefore beyond the RVB picture and may shed important light on the high-$T_c$ cuprate superconductors.}
\end{abstract}

\maketitle

\date{\today}


\noindent{\bf Introduction}\\

The Cooper pairing is the hallmark of superconductivity in both the conventional and the unconventional superconductors as evidenced by experiments like, e.g., the flux quantization. In the BCS theory, two electrons injected into a Fermi liquid can always form a Cooper pair under an arbitrarily weak attractive interaction. Ever since the discovery of the high-$T_c$ cuprates, a  great effort has been devoted to finding the responsible pairing glue, which is widely attributed to the superexchange interaction \cite{Anderson,Dagotto1994,White97,Chernyshev98,Pines_09,Scalapino12}.

However, at a deeper level, the BCS theory as a suitable description of the Cooper pairing has been seriously challenged in the cuprate \cite{Anderson07}. For instance, in a Mott insulator, the strong on-site Coulomb repulsion will cause the charge being stripped off the electrons, while their spins form the singlet RVB pairing in the RVB theory \cite{Anderson1987,lda88,Lee06}. Furthermore, in a doped Mott insulator, the original Fermion sign structure for a non-interacting Fermi gas is replaced by a much sparse sign structure (phase strings) \cite{Sheng1996,Weng1997,Wu-Weng-Zaanen,zaanen_09} as precisely identified in the $t$-$J$ model at arbitrary dimensions.

Without the integrity of individual electrons, to understand the nature of Cooper pairing in the doped cuprates, one has to go beyond the BCS scheme of simply identifying the pairing glue. Here, the behavior of the unpaired single-particle excitation has to be examined simultaneously. Recently, the loss of quasiparticle coherence in a Mott insulator has been studied for the single-hole doped $t$-$J$ square ladders by a large-scale DMRG simulation  \cite{ZZ2013}. Due to the destructive quantum interference effect of phase strings, a novel charge localization purely of strong correlation origin has been unveiled, which is independent of whether the underlying spin correlation is quasi-long-ranged (in odd-leg ladders) or short-ranged (in even-leg ladders).  It is also found that a coherent Bloch quasiparticle behavior can be recovered once the phase string effect is artificially turned off in the kinetic term of the $t$-$J$ model \cite{ZZ2013}.

Experimentally, the absence of a coherent quasiparticle excitation has been clearly observed by the angle-resolved photoemission spectroscopy (ARPES) in lightly doped cuprates such as Ca$_{2-x}$Na$_x$CuO$_2$Cl$_2$ \cite{ARPES98,ARPES04} as well as the underdoped YBa$_2$Cu$_3$O$_{6+x}$ \cite{ARPES10}. As a matter of fact, the single-particle excitation is generally frustrated in a normal state of the cuprate, from the antiferromagnetically (AF) ordered phase to the pseudogap regime, and to the strange-metal phase at the optimal doping \cite{shen_03}. The transport experiment has also universally shown the localization of charge carriers in the underdoped regime before superconductivity sets in \cite{Ando10}.

\begin{figure*}[tbp]
\begin{center}
\includegraphics[height=3.5in,width=6in]{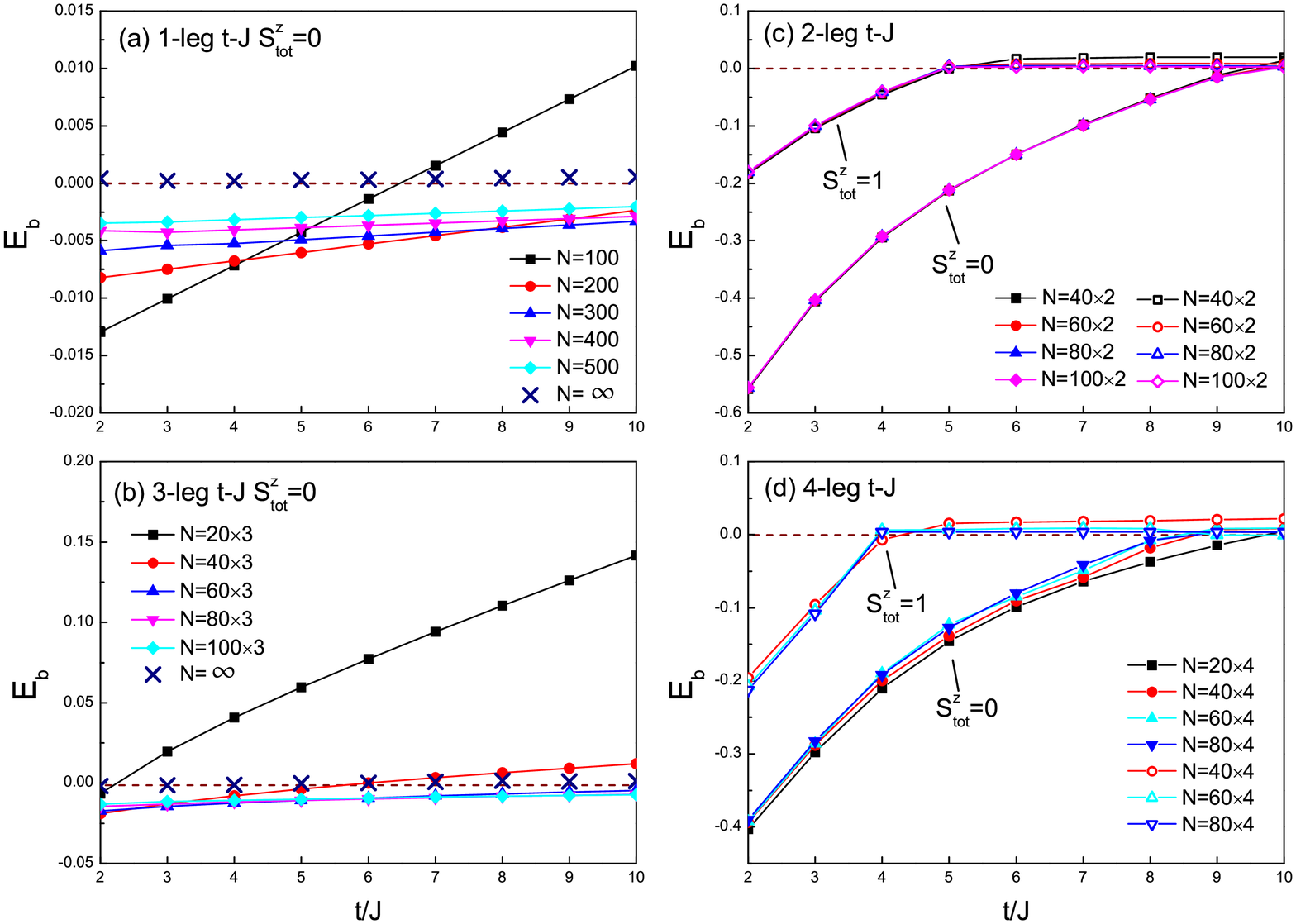}
\end{center}\textbf{}
\par
\renewcommand{\figurename}{Fig.}
\caption{(Color online) The binding energy $E_b$ for two holes in a $t$-$J$ ladder is obtained by the DMRG method, with the leg number ranging from $1$ in (a), $3$ in (b), and $2$ in (c), $4$ in (d).
For even-leg ladders with finite spin gaps, the binding strength is substantially larger than the odd-leg ladders without a spin gap. In the large length limit, the extrapolated binding energy almost vanishes for the odd-leg ladders, while remains big for the even-leg ladders until at much larger $t/J\gg 1$. It is interesting to note that for the even-leg ladders, the binding strength is strong not only in the singlet channel, but also in the triplet ($S^z_{tot}$=$1$) channel. }
\label{Eb}
\end{figure*}

In this paper, we further investigate the hole pairing in the $t$-$J$ ladders using the DMRG algorithm \cite{DMRG92}. A significantly strong binding energy is indeed found for two holes injected into a short-range-ordered  (even-leg) spin ladder, but the pairing strength becomes vanishingly small for two holes doped into a gapless (odd-leg) spin ladder. The origin of pairing here supports an RVB picture rather than the conventional BCS picture of exchanging the magnetic fluctuation. Namely, the hole pairing naturally arises from doping a short-ranged spin liquid state. Further surprisingly, the short-range spin correlation is only a necessary condition, but not a sufficient one. By turning off the phase string effect in the hopping term without changing the spin correlation, the coherent Bloch state is restored for the unpaired holes. But simultaneously the strong pairing disappears as well. It unveils a novel Cooper pairing mechanism, which works as an intrinsic combination of the spin RVB pairing with the charge pairing that removes the kinetic energy frustration of phase strings. Moreover, we verify that bound pairs of holes are generally repulsive to each other without further forming droplet. Our model study points to a non-BCS route to achieve high-$T_c$  in a doped Mott insulator, i.e., making the AF correlation a short-ranged one and, at the same time, frustrating the kinetic energy as much as possible by phase strings in the normal state.\\

\noindent{\bf Results}\\

{\bf Model Hamiltonians.} As a large-$U$ Hubbard model with the hopping integral $t\gg J$, where the superexchange coupling $J=4t^2/U$, the $t$-$J$ Hamiltonian is given by $H_{t\text{-}J} = H_t+H_J$ with
\begin{equation}
\begin{split}
H_t &= -t \sum_{\langle {ij}\rangle \sigma } {({c_{i\sigma }^{\dag}c_{j\sigma }+h.c.})}, \\
H_J &= J \sum_{\langle {ij}\rangle } {(\mathbf{S}_{i}\cdot \mathbf{S}_{j}-\frac{1}{4}n_{i}n_{j})},
\end{split}
\label{a}
\end{equation}
where $\langle ij\rangle $ stands for the nearest neighbors (NN). ${c_{i\sigma }^{\dagger }}$ is the electron creation operator at site $i$, ${\mathbf{S}_{i}}$ and ${n_{i}}$ are the spin and number operators, respectively. The Hilbert space is constrained by the no-double-occupancy condition, i.e., $n_{i}\leq 1$. Our study focuses on the ladders on square bipartite lattices of $N=N_x\times N_{y}$, where $N_{x}$ and $N_{y}$ are the site numbers in the $x$ and $y$ directions, respectively.  For the present DMRG simulation, we set $J=1$ as the unit of energy. \\

\begin{figure}[btp]
\begin{center}
\includegraphics[height=4in,width=3.2in]{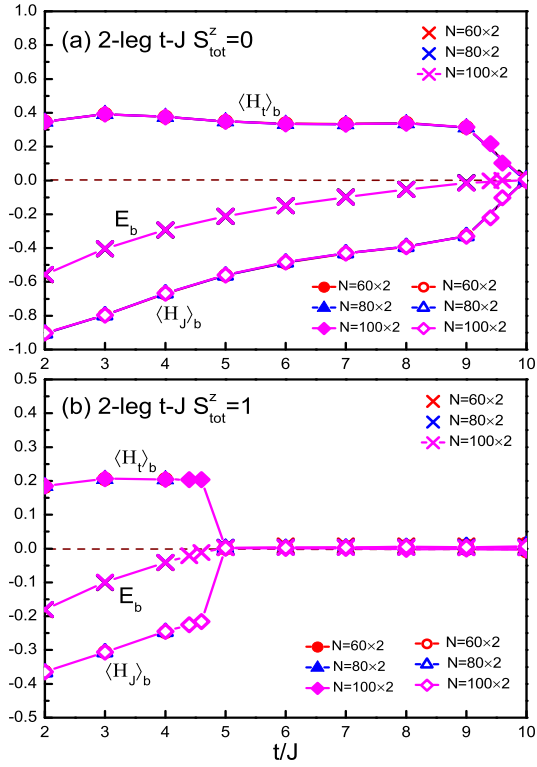}
\end{center}
\par
\renewcommand{\figurename}{Fig.}
\caption{(Color online)  The contributions of the hoping and superexchange terms to the binding energy are separated by ${E_b}$ = ${\left\langle {{H_t}} \right\rangle _b}$ + ${\left\langle {{H_J}} \right\rangle _b}$: (a) 2-leg ladder in the singlet ($S^z_{tot}=0$) channel. (b) 2-leg ladder in the triplet ($S^z_{tot}=1$) channel.  }
\label{Eb_separate}
\end{figure}

{\bf Binding energy.} Define the binding energy by
\begin{equation}
E_{b}=E_{2}+E_{0}-2E_{1},
\end{equation}
where  $E_2$  and $E_1$ are the ground-state energies of the two-hole and one-hole states, respectively, while $E_0$ denotes the ground-state energy at half filling. If two holes form a bound state, then $E_b$ is negative as $E_2-E_0$ $<$ $2(E_1-E_0)$. Otherwise, the binding energy $E_b$ should vanish in the thermodynamic limit with $E_2-E_0$ = $2(E_1-E_0)$ for two independent holes.

The binding energy $E_b$ is studied by the DMRG method in the two-hole doped $t$-$J$
ladders, with the leg number ranging from $N_{y} =1$ to 4 as shown in Fig.~\ref{Eb}. The binding strength is substantially large for an even-leg ladders [i.e., $N_y=2$ and $4$ in Figs.~\ref{Eb} (c) and (d), respectively], whereas $E_b$ diminishes quickly for an odd-leg ladder [i.e., $N_y=1$ and $3$ in Figs.~\ref{Eb} (a) and (b), respectively] with the increase of the ladder length $N_x$.  Indeed, for the odd-leg ladder cases, $E_b$ can be extrapolated to a vanishingly small value in the thermodynamic limit according to a finite size scaling using second-order polynomials of $1/N$ (see Supplementary Materials).

Note that the main distinction between the even- and odd-leg ladders is well known for the undoped case: there is a robust spin gap in the even-leg ladders with exponential-decay spin correlations at a length scale about 2 to 3 lattice spacing for $N_y=2$, $4$, but  the spin excitation is gapless for the odd-leg ladders with
quasi-long-range spin correlations \cite{White94,Dagotto1996}. It suggests that a \textit{short-range} AF correlation in the spin background should be a necessary condition for a meaningful pairing strength as found in Fig.~\ref{Eb}. Such an even-odd distinction in pairing has already been seen in the previous DMRG work at smaller lattice sizes \cite{White97}. As a matter of fact, it supports an RVB picture \cite{Anderson1987,Lee06} of pairing: two doped holes gain a binding energy by removing an RVB pair from the spin background. As for a weaker but still substantial binding energy for the triplet pairing, shown in Figs.~\ref{Eb} (c) and (d), it means that the lowest energy of a triplet spin excitation still lies below two \textit{free} spinon excitations in the even-leg ladders \cite{Tang13,Lake05}. By contrast, in the odd-leg ladders, the RVB pairing is long-ranged with gapless free spinon excitations, such that the binding energy is vanishingly small in the thermodynamic limit.

\begin{figure}[t]
\begin{center}
\includegraphics[height=2.2in,width=3.4in]{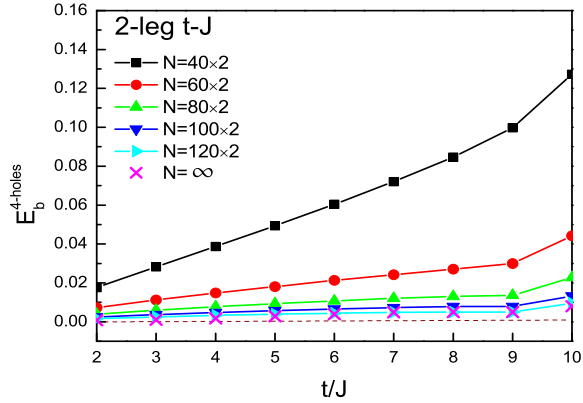}
\end{center}
\par
\renewcommand{\figurename}{Fig.}
\caption{(Color online) The binding energy of the two pairs in the four hole case, defined as $E_{b}^{\text{4-holes}}=(E_4-E_0)-2(E_2-E_0)$. It shows that two hole pairs are repulsive for all the finite systems, indicating that the binding exists only between two holes. }
\label{4holes}
\end{figure}

One can further distinguish the contributions from the kinetic and superexchange terms in the binding energy. For this purpose, we redefine ${E_b} = {\left\langle {{H_t}} \right\rangle _b} + {\left\langle {{H_J}} \right\rangle _b}$, where ${\left\langle {{H_t}} \right\rangle _b} \equiv {\left\langle {{H_t}} \right\rangle _{2}} - 2{\left\langle {{H_t}} \right\rangle _{1}}$ and ${\left\langle {{H_J}} \right\rangle _b} \equiv {\left\langle {{H_J}} \right\rangle _{2}} + {\left\langle {{H_J}} \right\rangle _{0}}-2{\left\langle {{H_J}} \right\rangle _{1}}$. Here $\left\langle {H_t} \right\rangle _2$ ($\left\langle {H_J} \right\rangle _2$) and $\left\langle {H_t} \right\rangle _1$ ($\left\langle {H_J} \right\rangle _1$) represent the kinetic energy (superexchange energy) of the two-hole and one-hole doped systems, respectively, and $\left\langle {H_J} \right\rangle _0$  the superexchange energy at half filling. In Figs.~\ref{Eb_separate} (a) and \ref{Eb_separate} (b), the separated contributions for the two-leg ladder are shown in the singlet and triplet channels, respectively. Clearly, the superexchange interaction serves as the driving force for the hole binding with ${\left\langle {{H_J}} \right\rangle _b}<0$ whereas ${\left\langle {{H_t}} \right\rangle _b} >0$. The hole binding eventually abruptly vanishes at $t/J>10$ (singlet pairing) and $t/J>5$ (triplet pairing) for the two-leg ladder in Fig.~\ref{Eb_separate}. Such abruptness is actually consistent with an RVB picture rather than a conventional pairing mechanism by exchanging magnetic fluctuations. In the latter, a smooth crossover to the disappearance of the Cooper pair is usually expected with the increase of $t/J$. Once the total binding energy equals to zero, one has ${\left\langle {{H_t}} \right\rangle _b} = {\left\langle {{H_J}} \right\rangle _b}=0$ such that the two doped holes behave independently. Note that at $t<J$, the binding energy $E_b$ is found to quickly reduce as the holes tend to stay at the two sides of the open boundary in the a DMRG calculation. But the boundary effect and phase separation at $t/J<1$ are no longer important once the kinetic energy becomes dominant over the superexchange energy at $t>J$, as illustrated in Figs.~\ref{Eb} and ~\ref{Eb_separate}, which is our main focus in this work.

Finally, we check that there is no formation of a ``droplet'' when more holes are added. Define the binding energy for two pairs of holes by $E_{b}^{\text{4-holes}} =(E_4-E_0)-2(E_2-E_0)$, where $E_4$ is the ground-state energy of the four-hole state. As shown in {Fig.}~\ref {4holes} for the two-leg $t$-$J$ ladder,  the two hole pairs are actually repulsive to each other at any finite size, and do not form a 4-hole droplet in the thermodynamic limit. Since each Cooper pair is well formed in a spin gapped state (its coherence will be further examined below), a superconducting condensation is naturally expected for a finite density of holes.\\

\begin{figure}[tbp]
\begin{center}
\includegraphics[height=5.85in,width=3.4in]{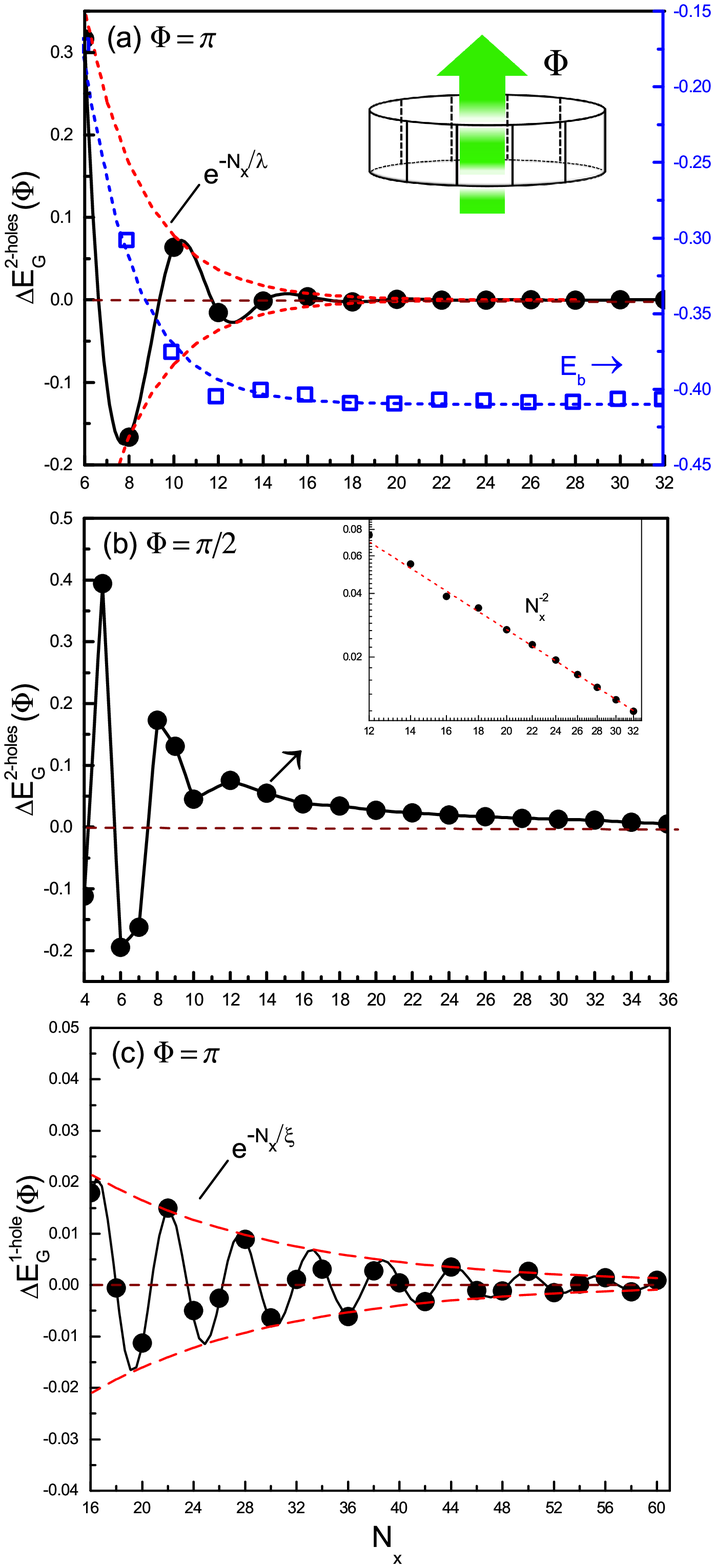}
\end{center}
\par
\renewcommand{\figurename}{Fig.}
\caption{(Color online) (a)  By inserting a flux $\Phi$ into a closed ribbon made of a two-leg ladder connected along the long chain direction (the inset), the energy difference between $\Phi=0$ and $\Phi\neq 0$: $\Delta E_{G}^{\text{2-holes}}(\Phi)\equiv E_2(\Phi)-E_2(0)$, is calculated at $t=3J$. Here $\Phi=\pi$ corresponds to the flux quantization condition, and $\Delta E_{G}^{\text{2-holes}}$ oscillates strongly and decays exponentially with a length scale $\lambda\approx$3, indicating the frustration of the phase string effect and the pairing size. The binding energy $E_b$ as a function of $N_x$ is also presented. (b) $\Delta E_{G}^{\text{2-holes}}$ at a non-quantized $\Phi=\pi/2$ exhibits a power-law decay at large $N_x$ (the inset), indicating that the centre-of-mass motion of the hole pair behaves like a phase-string-free coherent object, which is not shown in (a) because its contribution at $\Phi=\pi$ is the same as $\Phi=0$ (see the text). (c) For the single-hole case, $\Delta E_{G}^{\text{1-holes}}(\Phi)\equiv E_1(\Phi)-E_1(0)$  also exhibits an oscillation with an envelop of exponential decay, indicating \cite{ZZ2013} the self-localization of the hole with $\xi=14.5$ due to the phase string effect.}
\label{E_dif}
\end{figure}

{\bf Novel pairing mechanism.} In the following, we show that the presence of a spin gap/short-range spin correlation (i.e., the RVB mechanism) is only a necessary but not a sufficient condition for the appearance of strong binding between the holes. Instead, a new pairing mechanism hidden in the kinetic energy term of the $t$-$J$ model will play an essential role, which is of non-BCS type.

To examine the nature of pairing, we connect the ladder along the $x$ direction to make a close loop of circumference $N_x$ and then thread a flux $\Phi$ through the ring [see the inset of Fig.~\ref {E_dif} (a)]. Note that $\Phi$ here only couples to the doped holes in the hopping term via the usual U(1) degree of freedom of the conserved charge. It corresponds to the change of the boundary condition from a periodic one to anti-periodic one ($\Phi=\pi$) or a twisted boundary condition at a general flux $\Phi$ for the doped holes (but not the spins in the superexchange term).

Now we compute the ground-state energy difference
\begin{equation}\label{DeltaE}
\Delta E_{G}^{\text{2-holes}}(\Phi) \equiv E_2(\Phi )-E_2(0).
\end{equation}
If two holes are paired up, then when $N_x$ is larger than the pair size, one expects that  $\Delta E_{G}^{\text{2-holes}}(\Phi) $ quickly vanishes at $\Phi=\pi$, i.e., the flux-quantization condition, because a pair of holes will contribute to a $2\pi$ phase change by winding around the closed loop once. By contrast, for a general twisted boundary condition of $\Phi< \pi $, the contribution from a ``coherent'' Cooper pair of mass $M^*$ is expected to be $\Delta E_{G}^{\text{2-holes}}(\Phi) \propto (\Delta k_x)^2/(2M^*)$ with $\Delta k_x= \Phi/ N_x$.
\begin{figure}[t]
\begin{center}
\includegraphics[height=2.3in,width=3.2in]{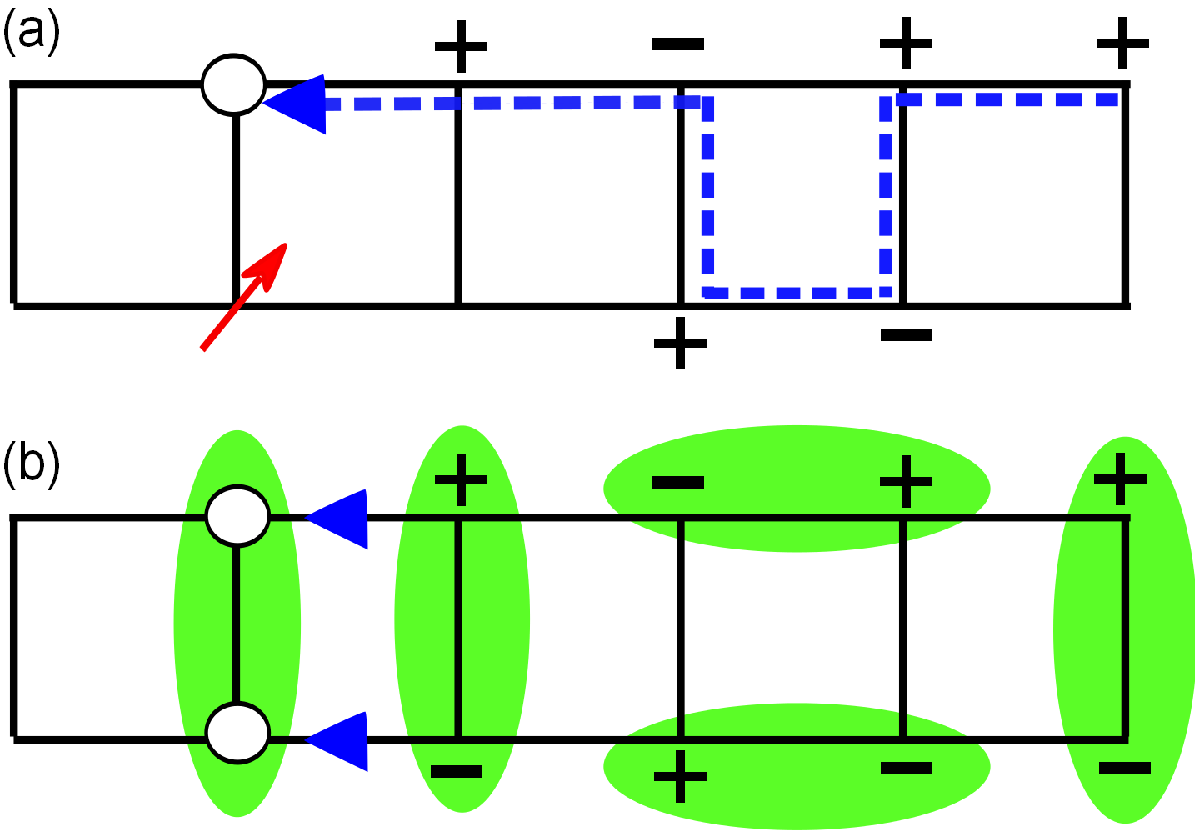}
\end{center}
\par
\renewcommand{\figurename}{Fig.}
\caption{(Color online) (a) The motion of a single hole (open circle) generally leaves a sequence of signs (i.e., the phase string) on its path, besides an unpaired spin (red arrow) in the spin background. (b) The binding of two holes can remove the unpaired spins to gain superexchange energy, but at the same time the frustration due to phase strings is compensated as well to gain the kinetic energy. Here a phase string is a product of the signs depending on the background spins exchanged with the hole during its hopping, and the short-range pairings of them effectively eliminate the destructive interference effect of phase strings.}
\label{phase string}
\end{figure}

Figures~\ref {E_dif} (a) and \ref {E_dif} (b) illustrate the behaviors of $\Delta E_{G}^{\text{2-holes}}$ for the two-leg ladder case at $\Phi=\pi$ and $\Phi=\pi/2$, respectively. In Fig.~\ref {E_dif} (a), with $\Phi$ satisfying the flux quantization condition $\Phi=\pi$, $\Delta E_{G}^{\text{2-holes}}$ oscillates strongly and falls off exponentially as $e^{-N_x/\lambda}$ with $\lambda \sim 3 $. $\lambda$ roughly measures the pairing size of two holes as indicated by $E_b$ shown in the same figure. On the other hand, in Fig.~\ref {E_dif} (b)  at $\Phi=\pi/2$, the coherent behavior of the hole pair as a whole is clearly shown, which follows an expected power-law decay $\propto N_x^{-2}$ at large $N_x$ [cf. the inset of Fig.~\ref {E_dif} (b)]. Similar behavior has been also checked for other $\Phi$'s deviating from the quantization $\pi$.

The above results confirm the hole pairing at a length scale comparable to the spin correlation length. Furthermore, the strong oscillation of $\Delta E_{G}^{\text{2-holes}}(\pi) $ at small sample sizes in Fig.~\ref {E_dif} (a) indicates a peculiar relative motion of a single hole within the bound pair. Note that previously a similar oscillation effect has been found \cite{ZZ2013} in $\Delta E_{G}^{\text{1-hole}}(\Phi) \equiv E_1(\Phi )-E_1(0)$ for a single hole doped case, which is replotted in Fig.~\ref {E_dif} (c) for comparison.

As discussed in Ref. \cite{ZZ2013}, such an oscillation in the single hole case [Fig.~\ref {E_dif} (c)] is a direct manifestation of the so-called phase string effect hidden in the $t$-$J$ model, which represents the non-perturbative quantum frustration introduced by hole hopping. Microscopically, the propagation of a single hole described by the $t$-$J$ model can be precisely expressed by a superposition of quantum amplitudes of all the paths, each carrying a unique sign sequence, $(+1)\times (-1)\times (-1)\times \cdots $ as illustrated in Fig.~\ref{phase string} (a) known as the phase string \cite{Sheng1996,Weng1997,Wu-Weng-Zaanen}. The sign $\pm $ in such a sequence keeps track of how the hole hops on the spin background by differentiating the microscopic processes of $\uparrow$- or $\downarrow$-spin exchanging with the hole at each step of hopping. The destructive interference of phase strings [Fig.~\ref{phase string} (a)], picked up by the hole from different paths, suppresses the forward scatterring and results in the localization of the hole based on the previous DMRG study \cite{ZZ2013} [cf. Fig.~\ref {E_dif} (c)].

In the presence of two holes, if they form a bound pair, the strong frustration on the kinetic energy caused by phase strings can be effectively removed [as schematically illustrated in Fig.~\ref{phase string}  (b) for the two-leg ladder]. In other words, the singular phase string effect in the $t$-$J$ model provides a new non-BCS pairing force in favor of the charge pairing. Indeed, the hole pair behaves like a coherence entity in Fig.~\ref {E_dif} (b) with the phase strings well cancelled out at large distance, which is obtained under a general twisted boundary condition with $\Phi\neq \pi \mod (\pi)$. As pointed out above, the residual phase string oscillation only shows up at the smaller $N_x$ in Fig.~\ref {E_dif} (a), where the finite-size behavior of $\Delta E_{G}^{\text{2-holes}}$ exhibits an uncompensated phase string effect within the hole pair.

As a matter of fact, such a phase string effect can be completely ``turned off'' if one replaces the hopping term $H_t$ in (\ref{a}) by
\begin{equation}
H_{\sigma \cdot t} = -t \sum_{\langle {ij}\rangle \sigma }\sigma {({c_{i\sigma}^{\dag }c_{j\sigma }+h.c.})},
\label{c}
\end{equation}
where an extra spin-dependent sign $\sigma=\pm$ is added, resulting in the so-called $\sigma$$\cdot$$t$-$J$ model \cite{ZZ2013}.

\begin{figure}[tbp]
\begin{center}
\includegraphics[height=4.2in,width=3.2in]{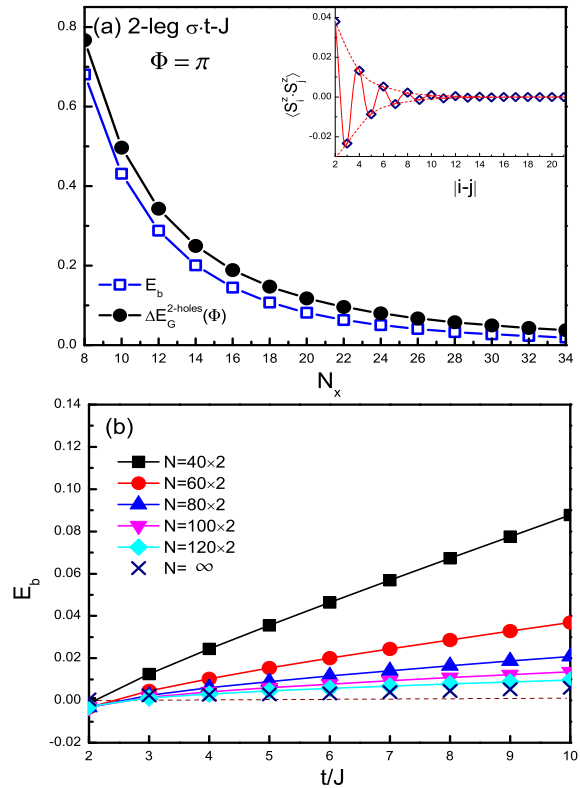}
\end{center}
\par
\renewcommand{\figurename}{Fig.}
\caption{(Color online) The hole binding in a two-leg ladder of the $\sigma$$\cdot$$t$-$J$ model, in which the phase string effect is switched off (see the text). (a)  The energy difference $\Delta E_{G}^{\text{2-holes}}(\Phi)$ at $\Phi=\pi$, is calculated at $t=3J$ for the geometry shown in the inset of Fig.~\ref {E_dif} (a). It decays in a power law fashion $\sim N_x^{-2}$, showing that a phase string free hopping term leads to a coherent quasiparticle behavior without binding. $E_b$ vanishes in the same power-law fashion. Note that spin correlations are still short-ranged as shown in the inset. (b) Binding energy for two holes as a function of $t/J$ at different sample sizes with the open boundary condition. The extrapolations to $N=\infty$ show that the strong binding found for the even-leg ladders in the $t$-$J$ model no longer exists here, and the binding strength becomes vanishingly weak. }
\label{NoPS_Eb}
\end{figure}

Then, for the $\sigma$$\cdot$$t$-$J$ model, the corresponding $\Delta E_{G}^{\text{2-holes}} $ exhibits no oscillation at all, which falls off in a power-law as $1 /N_{x}^{\alpha}$ with $\alpha = 2$ shown in {Fig.}~\ref {NoPS_Eb} (a) for the two-leg ladder. Such a power-law behavior simply implies that the holes propagate coherently as individual Bloch quasiparticles \cite{ZZ2013}, without pairing. The binding energy is also given in Fig.~\ref {NoPS_Eb} (a), which also decays in a power law fashion. In Fig.~\ref{NoPS_Eb} (b), $E_b$ as a function of $t/J$ is present at different ladder lengths, which is extrapolated to a vanishingly small value in the thermodynamic limit. Therefore, even with the same spin gap/short-range spin correlation  [cf. the inset of Fig.~\ref {NoPS_Eb} (a)] as in the $t$-$J$ case, by merely changing the hopping term to switch off the phase strings in (\ref{c}), a strong hole binding is significantly reduced to negligibly small. It means that the superexchange interaction is by no means the sole pairing glue in a doped Mott insulator.\\

\noindent{\bf Discussions}\\

In the literature it has been generally believed that the origin of Cooper pairing in a doped Mott insulator is solely due to the superexchange interaction. What has been unveiled in the present DMRG study is, surprisingly, that the hole pairing is actually achieved by a combination effect. That is, the superexchange and hopping terms in the $t$-$J$ model both play indispensable roles to the hole binding. On the one hand, the short-range spin correlation in a spin-gapped background is crucial to the spin RVB pairing. On the other hand, a strong frustration exerted on the kinetic energy of doped holes is also critical to force them to pair.

We have also studied the $\sigma$$\cdot$$t$-$J$ model. Its sole distinction compared with the $t$-$J$ lies in that phase string signs are completely ``switched off'', while the amplitude for each path remains the same as the absolute weight in the latter \cite{ZZ2013}. Here, with the constructive interference contributed by all the paths, a coherent quasiparticle behavior is restored for the unpaired hole propagation. Note that the spin superexchange correlation still remains the same, with the holes having the same tendency to pair up to gain superexchange energy. Nonetheless, the strong pairing found in the $t$-$J$ simply disappears, which unequivocally demonstrates that the non-BCS pairing force in the kinetic energy plays a critical role in the $t$-$J$ model.

So far what we have established is a novel pairing mechanism of a few doped holes in the $t$-$J$ ladders, in which the sample size ($N_y$) along the y-axis is deliberately kept small (up to $4$). The natural question is how much understanding that the present study provides is reliable on the real physics of the two-dimensional $t$-$J$ model at a finite doping, which is presumably relevant to the high-$T_c$ cuprate? Given the experimental facts that the Cooper pairing is more like a real space one in a spin background of short-range AF correlations, the lessons learned in the present model study may be highly valuable, especially with the spin and charge dynamics artificially adjustable through the leg numbers and by turning on/off the phase string effect. In particular, the phase string effect has been proven to be generally true for the $t$-$J$ model on any bipartite lattice \cite{Wu-Weng-Zaanen}. Therefore, the geometric limitation of the ladders in the present study is not expected to change the pairing mechanism fundamentally at a larger sample size (the leg number).

But there is one caveat. Namely, with the increase of the leg number, the spin gap in an undoped even-leg ladder should decrease monotonically, approaching to zero in the thermodynamic limit, where the AF long-range order is to be recovered. In the present study, however, we have found that a short-range spin correlation is crucial to the Cooper pairing. Thus the current results cannot be meaningfully extrapolated to the two-dimensional lattice at the same doping level. In other words, in the two-dimensional limit, a finite doping of holes will be needed in order to turn the gapless long-range-ordered AF state into a short-range-ordered paramagnet self-consistently. This is apparently beyond the scope of the present DMRG study. Nevertheless, our model study indicates that superconductivity has to arise in a short-range ``spin liquid'' background, with high-$T_c$ (large pairing strength) emerging out of a ``normal state'' where the unpaired holes get most severely frustrated, which is indeed consistent with the cuprate superconductivity.\\

\noindent{\bf Methods}\\

The numerical simulations in this work are performed by using the standard DMRG method \cite{DMRG92} on both the $t$-$J$ ladders in (\ref{a}) and $\sigma$$\cdot$$t$-$J$ ladders in (\ref{c}). Open boundary condition has been adopted in calculating the binding energies of the ladders, whose lengths have been extrapolated to the thermodynamic limit.  Periodic boundary condition as well as twisted boundary conditions, realized by threading fluxes into the closed loop made of the ladder as shown in the inset of Fig.~\ref{E_dif} (a), have been also used for probing the charge response. In the present DMRG simulations, we keep up to $m=5000$ states in the DMRG block with around 20-40 sweeps to get converged results. The truncation error is of the order or less than $10^{-8}$.

\vspace{12 pt}

\noindent{\bf Acknowledgements}\\

Stimulating and useful discussions with C.-S. Tian, Y. Qi, L.Balents and S.-S. Gong are acknowledged. This work was supported by the NBRPC Grant no. 2010CB923003, the NSF Grants DMR-0906816 and  PREM DMR-1205734 (DNS).\\

\noindent{\bf Author Information}\\

\noindent{\bf Affiliations}\\
\noindent{Institute for Advanced Study, and Collaborative Innovation Center of Quantum Matter, Tsinghua University, Beijing, 100084, China}\\
\noindent Zheng Zhu $\&$ Zheng-Yu Weng\\
\noindent{Department of Physics, University of California, Berkeley, 94720, USA}\\
\noindent Hong-Chen Jiang\\
\noindent{Department of Physics and Astronomy, California State University, Northridge, CA, 91330, USA}\\
\noindent D. N. Sheng\\

\noindent{\bf Supplementary Information:}\\

\begin{figure}[bp]
\begin{center}
\includegraphics[height=2.2in,width=6in]{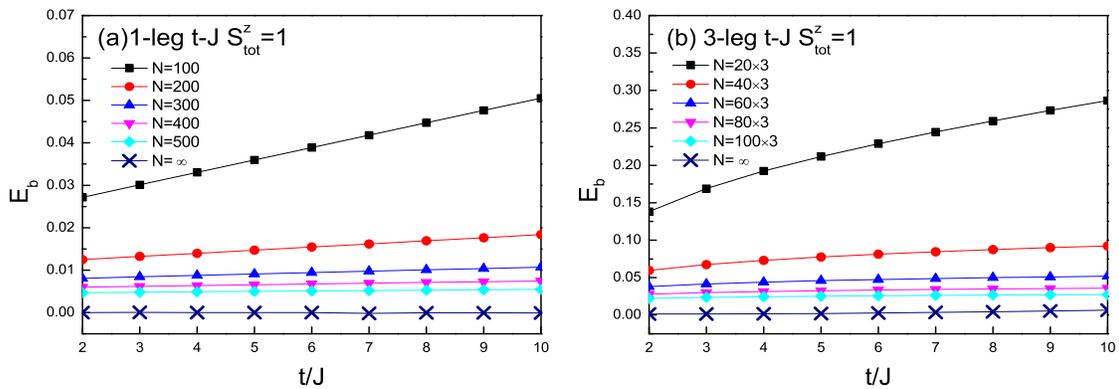}
\end{center}
\par
\renewcommand{\figurename}{Fig.}
\caption{(Color online) Two-hole binding energy $E_b$ in a triplet ($S^z_{tot}=1$) state :  (a) 1-leg $t$-$J$ chain; (b) 3-leg $t$-$J$ ladder. The binding energy in the thermodynamic limit with $N_x\rightarrow \infty$ is vanishingly small according to the finite size scaling using second-order polynomials of $1/N$.}
\label{Eb_Sz1}
\end{figure}

In this supplemental material, we present some technical details. As an unbiased and powerful numerical tool to study strongly correlated systems in one dimensional and quasi-one dimensional ladder systems,  the DMRG method is not limited by the exponential difficulty of total sites in exact diagonalization (ED) or the sign problem in quantum Monte Carlo (QMC) simulations. Since the ladders, serving as a bridge from one dimensional chains to two-dimensional layered systems, can be handled numerically to large sizes, they offer a chance to study the interplay of charge and spin degrees of freedom in the strongly correlated $t$-$J$ model. Moreover, a Heisenberg ladder with even number of legs has a short-ranged spin correlation and a finite energy gap in magnetic excitations,\cite{Dagotto96} and the corresponding ground state may be regarded as the short-ranged RVB state proposed by Anderson,\cite{Anderson87}. It may then lead to the hole pairing and superconductivity when holes are injected into the system as envisaged in Ref. \cite{Anderson87}.

The binding energy of the two-hole-doped  $t$-$J$ model is defined as $E_{b}=(E_{2}-E_{0})-2(E_{1}-E_{0})$. For the present DMRG simulation, doping two holes is realized by taking two electrons out of the half-filling, and the state of the total spins can be either singlet with $S^z_{tot}=0$ (removing an up spin and a down spin) or triplet with $S^z_{tot}=1$ (removing two down spins). For the odd-leg ladders, we have shown the results of the singlet state in the main text, where one can see that there is no strong binding when $S^z_{tot}=0$. Here in Fig.~\ref{Eb_Sz1}, the binding energies of the 1-leg and 3-leg $t$-$J$ ladders in the triplet channel are shown. The binding energy is vanishingly small for different values of $t/J$ in the thermodynamic limit, where the spin excitations behave like free spinons with a power-law decay of the spin-spin correlation length.

\begin{figure}[tbp]
\begin{center}
\includegraphics[height=3.9in,width=6.2in]{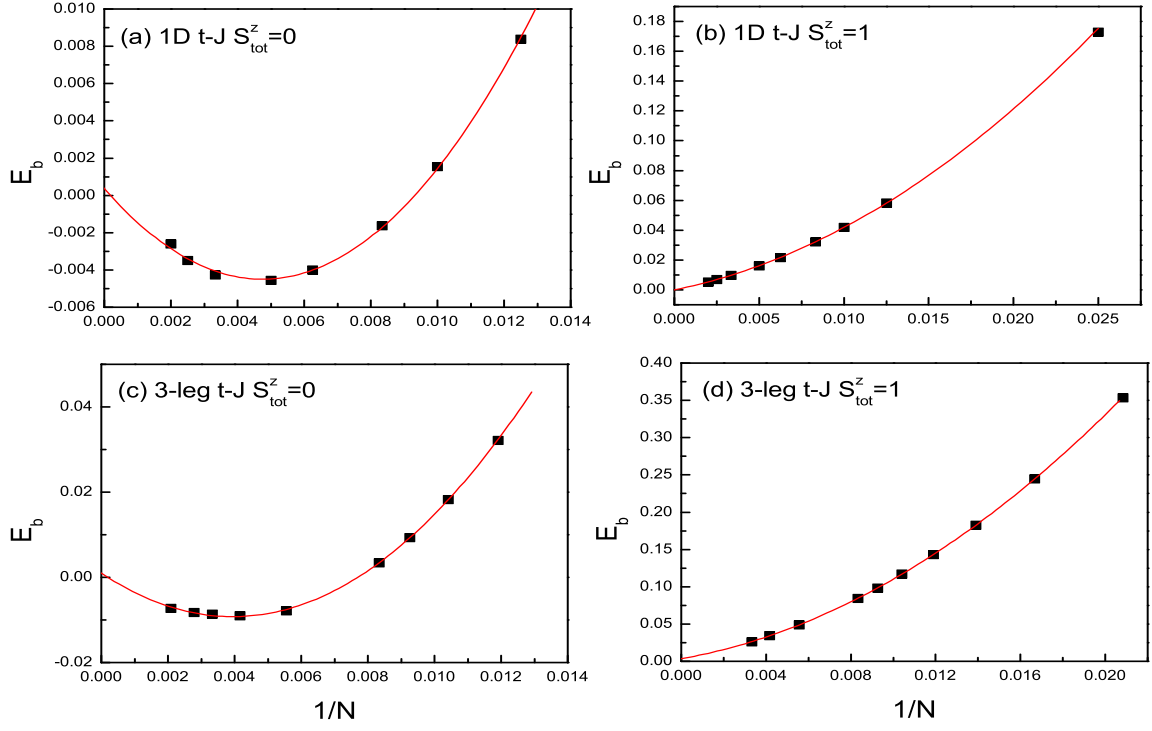}
\end{center}
\par
\renewcommand{\figurename}{Fig.}
\caption{(Color online) The extrapolation of the binding energy for the two-hole-doped $t$-$J$ ladders at $t/J=7$: (a)
1-leg with $S^z_{tot}=0$; (b) 1-leg with $S^z_{tot}=1$; (c) 3-leg with $S^z_{tot}=0$; and (d) 3-leg with $S^z_{tot}=1$. We extrapolate the results to the thermodynamic limit according to a general form $E_{b} (N)=E_{b} (N=\infty)+a/N+b/N^2$. The results show that the binding energy is almost zero when $N\rightarrow\infty$.}
\label{scaling_odd}
\end{figure}

\begin{figure}[tbp]
\begin{center}
\includegraphics[height=2.4in,width=6.2in]{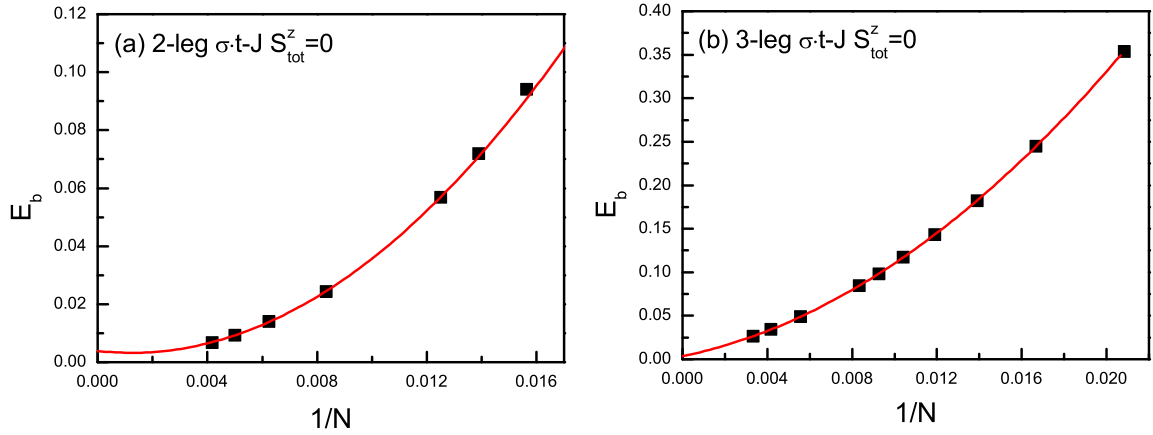}
\end{center}
\par
\renewcommand{\figurename}{Fig.}
\caption{(Color online) Typical examples of the extrapolation for the binding energy of two-hole-doped $\sigma\cdot t$-$J$ ladders at $t/J=7$: (a) 2-leg with $S^z_{tot}=0$; (b) 3-leg with $S^z_{tot}=0$. The extrapolation is made by using a general form $E_{b} (N)=E_{b} (N=\infty)+a/N+b/N^2$. The binding energy disappears at $N\rightarrow \infty$.}
\label{scaling_sigma}
\end{figure}

\begin{figure}[tbp]
\begin{center}
\includegraphics[height=2.2in,width=3.6in]{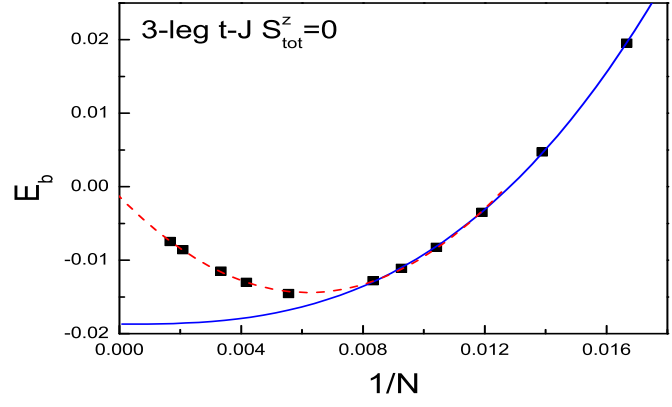}
\end{center}
\par
\renewcommand{\figurename}{Fig.}
\caption{(Color online) Typical examples of the extrapolation of the binding energy for two-hole-doped 3-leg $t$-$J$ ladders at $t/J=3$ with $S^z_{tot}=0$. The blue solid line
shows that if one only uses the data of smaller sizes for a finite-size scaling, it could lead to a wrong conclusion at $N_x\rightarrow \infty$. The correct extrapolation according to $E_{b} (N)=E_{b} (N=\infty)+a/N+b/N^2$, as presented by the red dashed line, shows that the binding energy is actually vanishingly small at $N=\infty$.}
\label{scaling_compare}
\end{figure}

Based on the results of the binding energy $E_b$ calculated at finite-length ladders, we can extrapolate the data to thermodynamic limit (i.e., $N_x=\infty$) according to a general fitting form $E_{b} (N)=E_{b} (N=\infty)+\frac{a}{N}+\frac{b}{N^2}$. Figure ~\ref{scaling_odd} shows the extrapolation of the binding energy for the two-hole-doped $t$-$J$ ladders (with $N_y=1$, $3$) in both singlet channel ($S^z_{tot}=0$) and triplet channel ($S^z_{tot}=1$) at $t/J=7$, which approaches vanishingly small binding energy at $N_x\rightarrow \infty$. Similarly, Fig.~\ref{scaling_sigma} shows the extrapolation of the binding energy for two-hole-doped $\sigma\cdot t$-$J$ ladders ($N_y=2$, $3$) with $S^z_{tot}=0$ at $t/J=7$, which also indicates that no bound state exists in the thermodynamic limit. Figure ~\ref{scaling_compare} further compares the contrasting results of two different extrapolations, which illustrates the importance to calculate sufficiently long ladders in order to capture the right results in the thermodynamic limit, indicating that a finite-size scaling based on small clusters may lead to wrong extrapolations.

\end{document}